\documentclass{iau}

\usepackage{amsmath}
\usepackage{graphicx}
\usepackage{multirow}

\newcommand{\yr}{{~\rm yr}}

\begin{document}

\lefttitle{N. Soker}
\righttitle{Violent mass ejection by the brightest PNe}

\journaltitle{Planetary Nebulae: a Universal Toolbox in the Era of Precision Astrophysics}
%\jnlPage{1}{7}
\jnlDoiYr{2023}
\doival{10.1017/xxxxx}
\volno{384}

\aopheadtitle{Proceedings IAU Symposium}
\editors{O. De Marco, A. Zijlstra, R. Szczerba, eds.}
 
\title{Violent mass ejection by the progenitors of the brightest planetary nebulae: supernova progenitors
}

\author{Noam Soker}
\affiliation{Department of Physics, Technion, Haifa, 3200003, Israel; soker@physics.technion.ac.il}

\begin{abstract}
I examine the morphologies of the brightest planetary nebulae (PNe) in the Milky Way Galaxy and conclude that violent binary interaction processes eject the main nebulae of the brightest PNe. The typical morphologies of the brightest PNe are multipolar, namely have been shaped by two or more major jet-launching episodes at varying directions, and possess small to medium departures from pure point symmetry. I discuss some scenarios, including a rapid onset of a common envelope interaction and the merger of the companion, mainly a white dwarf, with the asymptotic giant branch core at the termination of the common envelope. Some of these might be progenitors of type Ia supernovae (SNe Ia), as I suggest for SNR G1.9+0.3, the youngest SN Ia in the Galaxy.  
\end{abstract}


\maketitle

% ====================================
\section{Introduction}
\label{sec:Introduction}
% ====================================

The planetary nebula (PN) luminosity function (PNLF) has a more or less universal cutoff at the bright end (e.g., \citealt{Ciardulloetal1989, Jacoby1989, vandeSteeneetal2006, Ciardullo2022, Chornay2023PhD, Jacobyetal2023}), although variations exist (e.g., \citealt{Bhattacharyaetal2021}). In this study, I do not attempt to explain this cutoff as it requires the calculations of the evolution of the central star of the PN and radiative transfer (e.g., \citealt{Mendez2017, Gesickietal2018, SabachSoker2018, Valenzuelaetal2019}). I limit myself to examining the morphologies of the brightest PNe in the Milky Way Galaxy. 

I examine some morphological features that are much more common in the brightest Milky Way PNe (Section \ref{sec:features}) than in the hundreds of well resolved PNe that different surveys and catalogs present (e.g.,   \citealt{Balick1987, Chuetal1987, Schwarzetal1992, CorradiSchwarz1995, Manchadoetal1996, SahaiTrauger1998, Sahaietal2011, Parkeretal2016, Parker2022}). Many PNe and proto-PNe with points-symmetric morphologies and/or with pairs of opposite lobes were likely shaped by jets (e.g., \citealt{Morris1987, Soker1990AJ, SahaiTrauger1998, Sahai2000, Sahaietal2000, Sahaietal2007, Sahaietal2011, Boffinetal2012}).  

Based on the shaping of point-symmetric morphologies by jets, I argue in Section \ref{sec:G1903} that the newly identified point-symmetry in the youngest supernova in the Galaxy implies that this type Ia supernova (SN Ia) exploded inside a PN (\citealt{Soker2023SNRG1903}).

In Section \ref{sec:Summary}, I summarize this short study by speculating that many of the brightest PNe are the results of interaction with a WD companion. A small fraction of these PNe eventually explodes as SNe Ia. 

% ====================================
\section{Morphological features of the brightest Galactic PNe}
\label{sec:features}
% ====================================

\cite{Chornay2023Pr} lists extinction corrected bright PNe in [O III] (M(5007); for an earlier list see \citealt{Mendezetal1993}). The 13 brightest PNe in his list are as follows (from brightest to dimmest). NGC 6572; IC 4670 (Hb 6); K 3-17; M 1-40; NGC 7027; NGC 6369; NGC 7354; NGC 7662; NGC 6537; NGC 7026; NGC 6567; NGC 2440; NGC 6210. 
I take the 13 brightest PNe and not more because of limited space in this Proceedings paper.  

\begin{figure}
\begin{center}
\includegraphics[trim=0.0cm 17.3cm 0.0cm 1.4cm ,clip, scale=0.64]{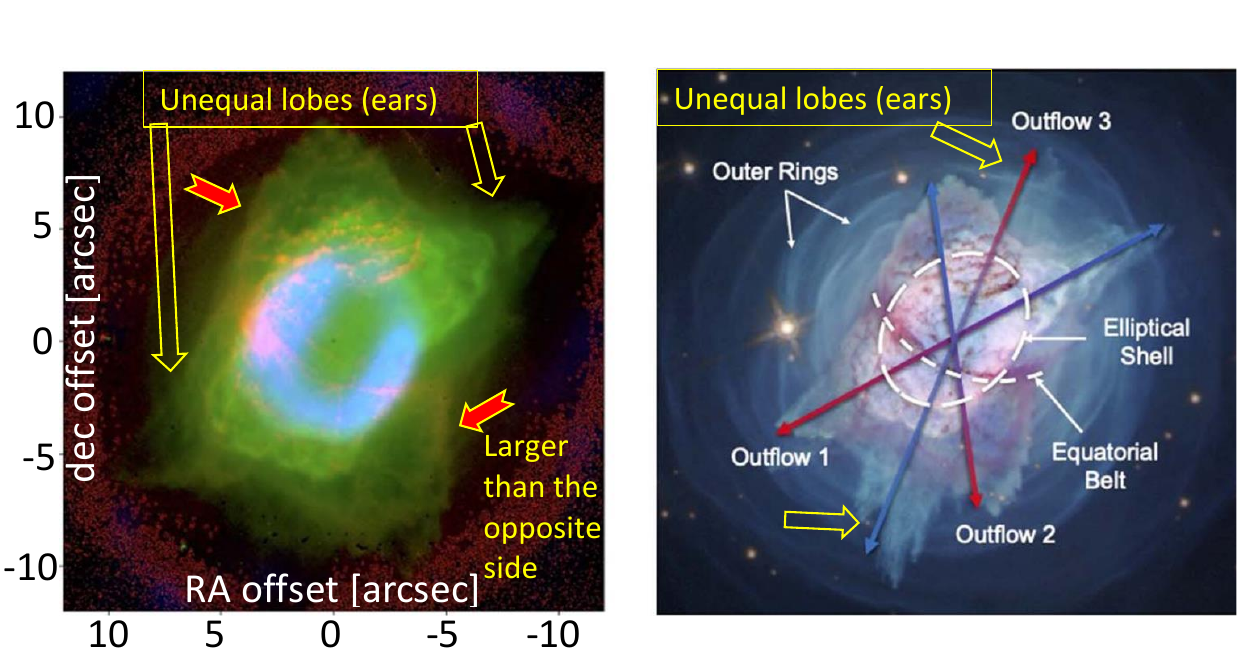}
\end{center}
\caption{Two panels from \cite{MoragaBaezetal2023} (their figure 10) of NGC 7027. On the left the red color depicts the Pa$\beta$/H$\beta$ line ratio, the green is extinction-corrected H$\alpha$ emission, and the blue is 1.3 mm radio continuum from \cite{Bublitzetal2023}. On the right panel, \cite{MoragaBaezetal2023} schematically mark some structural components. They mark the three pairs of jets as Outflow 1, 2, and 3, where red/blue colors indicate the
redshift/blueshift. Also marked are the nebula’s outer ring system, central elliptical shell, and equatorial belt. I added double-lined yellow arrows that point to morphological features that depart from pure point symmetry, as the insets indicate. 
}
\label{fig:NGC7027} 
\end{figure}

I start with NGC 7027, one of the brightest PNe in the Galaxy, and refer to the recent analysis of this PN by \cite{MoragaBaezetal2023}. In this thorough analysis \cite{MoragaBaezetal2023} present images in different lines and line ratios. In Figure \ref{fig:NGC7027} I present two of their images. Several clear morphological features deviate from pure point symmetry. For example, the two opposite lobes (ears) in all outflows are not equal to each other (I mark two pairs). I mark a pair of opposite bright edges (yellow-red arrows). These are not equal in size, nor do they have the same distance from the center.     
% FFFFFFFFFFFFFFFFFFFFFFFFFFFFFFFFFFFFFFFFFFFFFFFFFFFFFF
% FFFFFFFFFFFFFFFFFFFFFFFFFFFFFFFFFFFFFFFFFFFFFFFFFFFFFFFF

The image of NGC 7027 presents the following properties, which are also seen in many earlier studies of NGC 7027 (e.g., \citealt{Bublitzetal2023}). 
\newline 
\textbf{(1)} \textit{Multipolar. }It is a multipolar PN, with three outflow directions (e.g., \citealt{Coxetal2002}). 
\newline
\textbf{(2)} \textit{Departure from symmetry.} There is a departure from pure point symmetry in the sense that opposite clumps/lobes/arcs/filaments are not equal in one or more of their properties, including differences in brightness, size, structure, distance from the center, and/or bending such that the two clumps are not exactly opposite to each other through the center.  
\newline
\textbf{(3)} \textit{Not messy.} Despite these departures from pure point-symmetry, the structure is not `messy'. Namely, all morphological features present some symmetry and there are no strong and large morphological features that lack any symmetry. Messy PNe are most likely to result from triple-star interaction (e.g., \citealt{BearSoker2017}). NGC 7027 does not require a tertiary star to explain its basic morphology.

\begin{figure}
\begin{center}
\includegraphics[trim=0.0cm 14.0cm 0.0cm 0.0cm ,clip, scale=0.55]{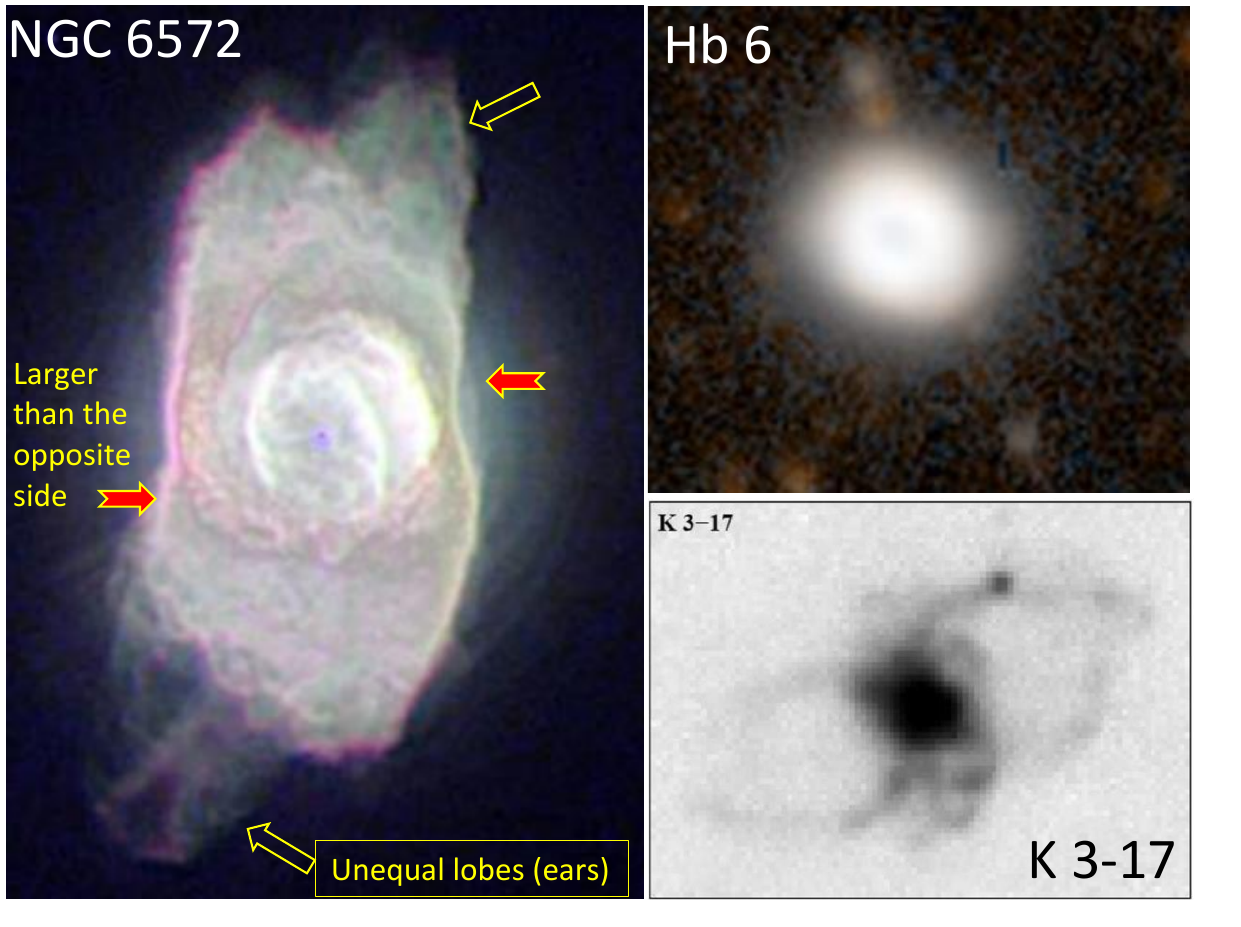}
\end{center}
\caption{Left: An HST image of NGC 6572 in H$\alpha$ (red), H$\beta$ (dark blue), oxygen (blue), and a yellow broadband filter (green) (credit:
ESA/Hubble \& NASA). I indicated departures from point symmetry. For more details on the multipolar structure see, e.g., \cite{Bandyopadhyayetal2023}.
The marks emphasize the three properties of (1) multipolar morphology, (2) departure from symmetry, and (3) non-messy morphology (only a small to moderate departure from pure point symmetry).
Upper right: an HST image of IC 4670 (Hb 6; \citealt{Chambersetal2016}). Lower right: 
K3-17 (from \citealt{Steffenetal2013}).  }
\label{fig:NGC6572} 
\end{figure}

The brightest PN is NGC 6572 (Figure \ref{fig:NGC6572}). A recent study by \cite{Bandyopadhyayetal2023} reveals its 3D structure with morphokinematic and photoionization modeling. It is also multipolar (2 pairs of lobes) with a clear departure from pure point symmetry, but it is not a messy PN. Namely, it shares the same three properties that I listed above for NGC 7027. The image of IC 4670 (Hb 6) that I take from \cite{Chambersetal2016} is of too low a resolution to examine the three properties in detail. However, the jet-like structures above and below the main ring in the image suggest that it might share these three properties. 
In the same figure I also present K3-17 taken from \cite{Steffenetal2013} who studied this PN as a multipolar PN. It also shares the three properties that I listed for NGC 7027.    
% FFFFFFFFFFFFFFFFFFFFFFFFFFFFFFFFFFFFFFFFFFFFFFFFFFFFFF
% FFFFFFFFFFFFFFFFFFFFFFFFFFFFFFFFFFFFFFFFFFFFFFFFFFFFFFFF

\begin{figure}
\begin{center}
\includegraphics[trim=0.0cm 16.0cm 0.0cm 0.0cm ,clip, scale=0.66]{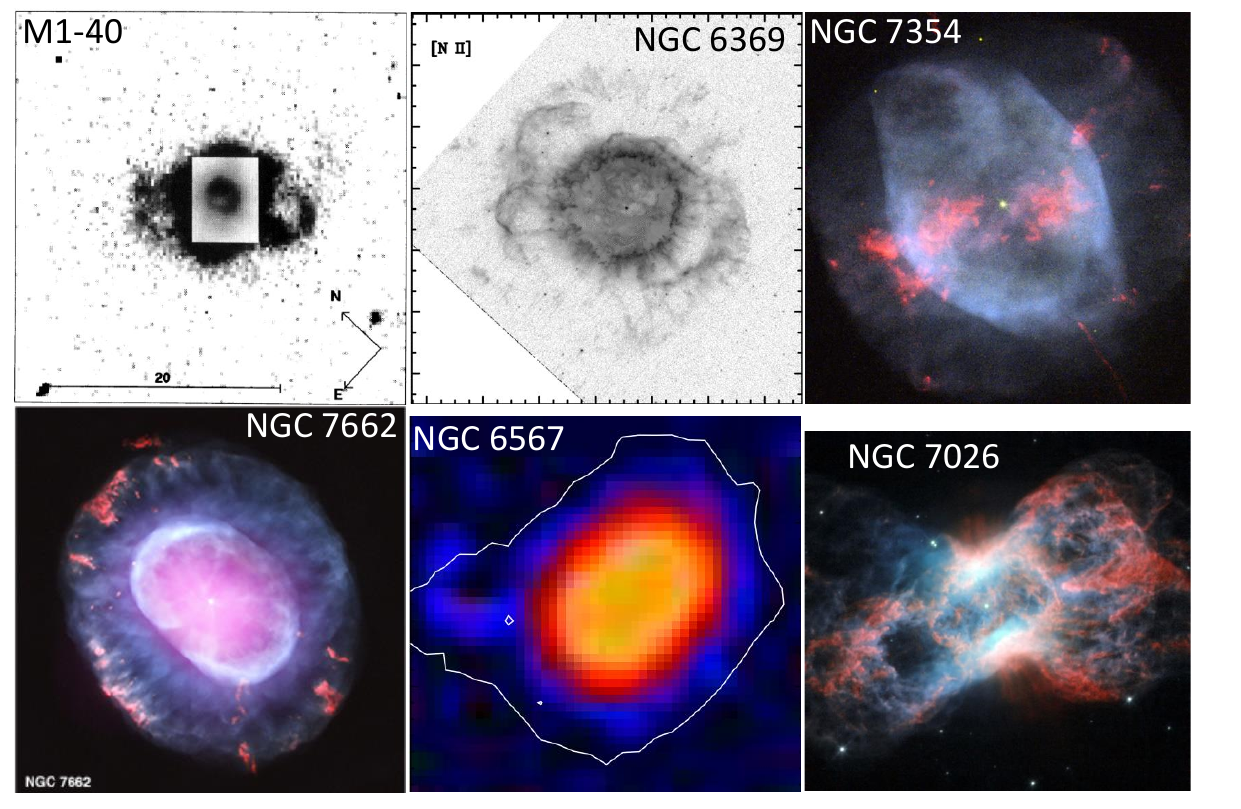}
\end{center}
\caption{Six bright PNe, from upper left to lower right. An [O \textsc{iii}] image of M1-40 (\citealt{Schwarzetal1992}), an HST [N \textsc{ii}] image of NGC 6369 \citep{RamosLariosetal2012}, an image of NGC 7354 from HST site (where more details can be found; credit ESA/Hubble \& NASA), NGC 7662 (from the Chandra site based on \citealt{Kastneretal2012}), an HST [N \textsc{ii}] (log flux scale) image of NGC 6567 from \cite{Danehkar2022}, and an HST image of NGC 7026 from the HST site (credit ESA/Hubble \& NASA).  
}
\label{fig:6PNe} 
\end{figure}

In Figure \ref{fig:6PNe} I present the images of six bright PNe. 
The PN NGC 7026 has two pairs of lobes and has a departure from pure point symmetry as the structural analysis by \cite{Clarketal2013} shows (lower-right panel of Figure \ref{fig:6PNe}). As with some other PNe above, it does not have a strong messy morphology. 
The image of the PN NGC 7354 in the upper-right panel of Figure \ref{fig:6PNe} presents only one symmetry axis (from upper left to lower right) and clumps near the equatorial plane (in red; perpendicular to the main symmetry axis). However, deeper analysis by \cite{Contrerasetal2010} reveals two axisymmetrical axes inclined to each other. It seems that NGC 7354 also shares the properties I listed for NGC 7027, although only with a weak departure from axisymmetry. NGC 6369 shows two lobes on one side (left in the upper-middle panel). Clearly, its morphology departs from pure point symmetry and it seems to be multipolar, sharing the three properties with the other PNe mentioned above. The PN M1-40 might also share the three properties, although a better-resolution image is required. The case with NGC 6567 is unclear, while NGC 7662 does not reveal a multipolar structure.  
% FFFFFFFFFFFFFFFFFFFFFFFFFFFFFFFFFFFFFFFFFFFFFFFFFFFFFF
% FFFFFFFFFFFFFFFFFFFFFFFFFFFFFFFFFFFFFFFFFFFFFFFFFFFFFFFF

\begin{figure}
\begin{center}
\includegraphics[trim=0.0cm 9.0cm 0.0cm 0.0cm ,clip, scale=0.48]{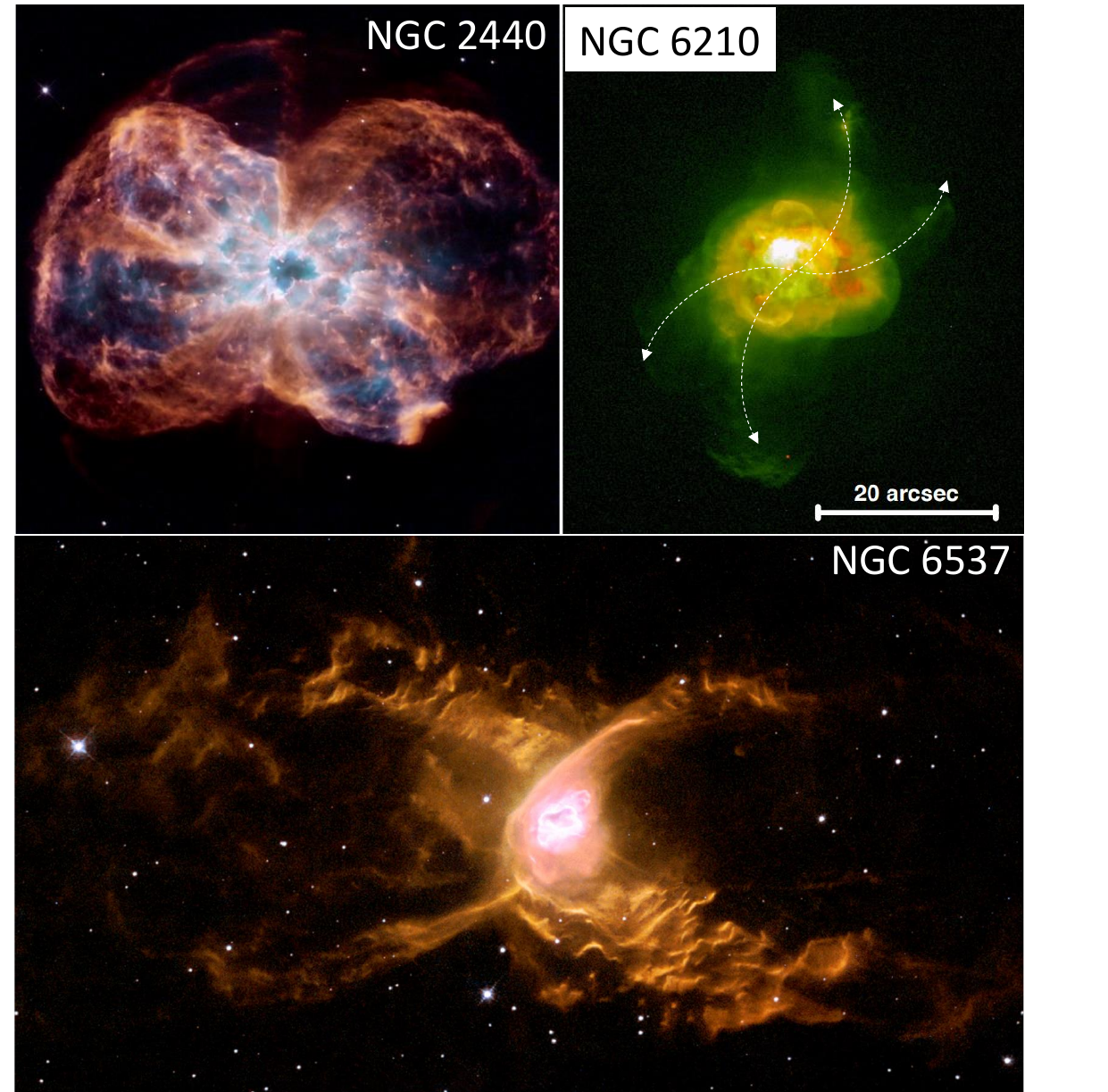}
\end{center}
\caption{Upper left: NGC 2440 (The HST site where more details can be found; credit: NASA, ESA, and K. Noll, STScI). Upper right: An [O \textsc{iii}] image in log scale of the PN NGC 6210 (\citealt{Henneyetal2021}). I added two S-shaped double-headed arrows to mark the S-shaped pairs of lobes.    
Bottom: an HST image of NGC 6537 (credit: SA \& Garrelt Mellema). It has a large-scale bipolar morphology, but with clear departures from pure point symmetry and a multipolar morphology in the inner (white) region.
}
\label{fig:3PNe} 
\end{figure}

The PN NGC 2440 has two pairs of bipolar lobes as the morpho-kinematical modeling by \cite{Lagoetal2016} shows, and as can also be seen in the upper left panel of Figure \ref{fig:3PNe}. 
A departure from pure point symmetry is seen. NGC 2440 also shares the three properties that I listed for NGC 7027. \cite{BearSoker2017} mark NGC 2440  as likely (moderate probability) shaped by a triple stellar system. 
The PN NGC 6210 is a multipolar PN (e.g., \citealt{RechyGarciaetal2020, Henneyetal2021}). It is different than most PNe in this study in that \cite{BearSoker2017} classified it as high-probability messy PN. In that case, it requires a triple star system interaction to explain its morphology (\citealt{BearSoker2017}). However, its outer structure might be viewed as composed of two S-shaped pairs of lobes (as I marked in the figure), which would then make NGC 6210 morphology less dominated by the messy parts. The morphology of NGC 6537 is different than those of the other PNe in this study but shares the basic properties. On a large scale, it is a bipolar PN with two opposite large lobes. The boundaries of the lobes are not smooth, and the two lobes are not exactly equal to each other in shape. In the inner region (white) there is a hint of a multipolar PN, although on a smaller scale than the large lobes. The inner regions depart from pure-point symmetry. Although it is different than the other PNe, it is a multipolar PN with a departure from pure point symmetry, but not a messy PN. 
% FFFFFFFFFFFFFFFFFFFFFFFFFFFFFFFFFFFFFFFFFFFFFFFF
% FFFFFFFFFFFFFFFFFFFFFFFFFFFFFFFFFFFFFFFFFFFFFFFFFFFFFFFF

Overall, I find that most, but not all, bright PNe tend to be multipolar with small to medium degree departures from pure point-symmetry. These properties suggest that they were shaped by a strong (violent) binary interaction, but not necessarily by a triple-star interaction that most likely leads to messy PNe. On the opposite extreme, there are faint PNe that tend to be round (e.g., \citealt{Jacobyetal2010}), most likely because they did not have a binary interaction at all or had an interaction with a sub-stellar object (e.g., \citealt{SokerSubag2005}). 
Out of the 13 PNe I study here, \cite{BearSoker2017} attributed a high probability for shaping by triple stellar interaction only to NGC 6210 (but see the marks for more ordered S-shaped pairs of lobes in Figure \ref{fig:3PNe}), and a medium probability for NGC 2440. But there are no indications of a large influence by a tertiary star in the other bright PNe. The binary interaction, though, seems to require the change of the symmetry axis of the jets within a relatively short time, during which the system launches two or more energetic pairs of jets (this is what I refer to as `violent interaction'). 

% ====================================
\section{The case of SN Ia SNR G1.9+0.3 that destroyed its planetary nebula progenitor}
\label{sec:G1903}
% ====================================

\begin{figure*}[]
	\centering
%	\hspace*{-2cm} 
	% This cut edges: [trim=left bottom right top, clip]{file}
%	\hspace{1cm}
\includegraphics[trim=1.2cm 8.5cm 2.5cm 2.0cm ,clip, scale=0.4]{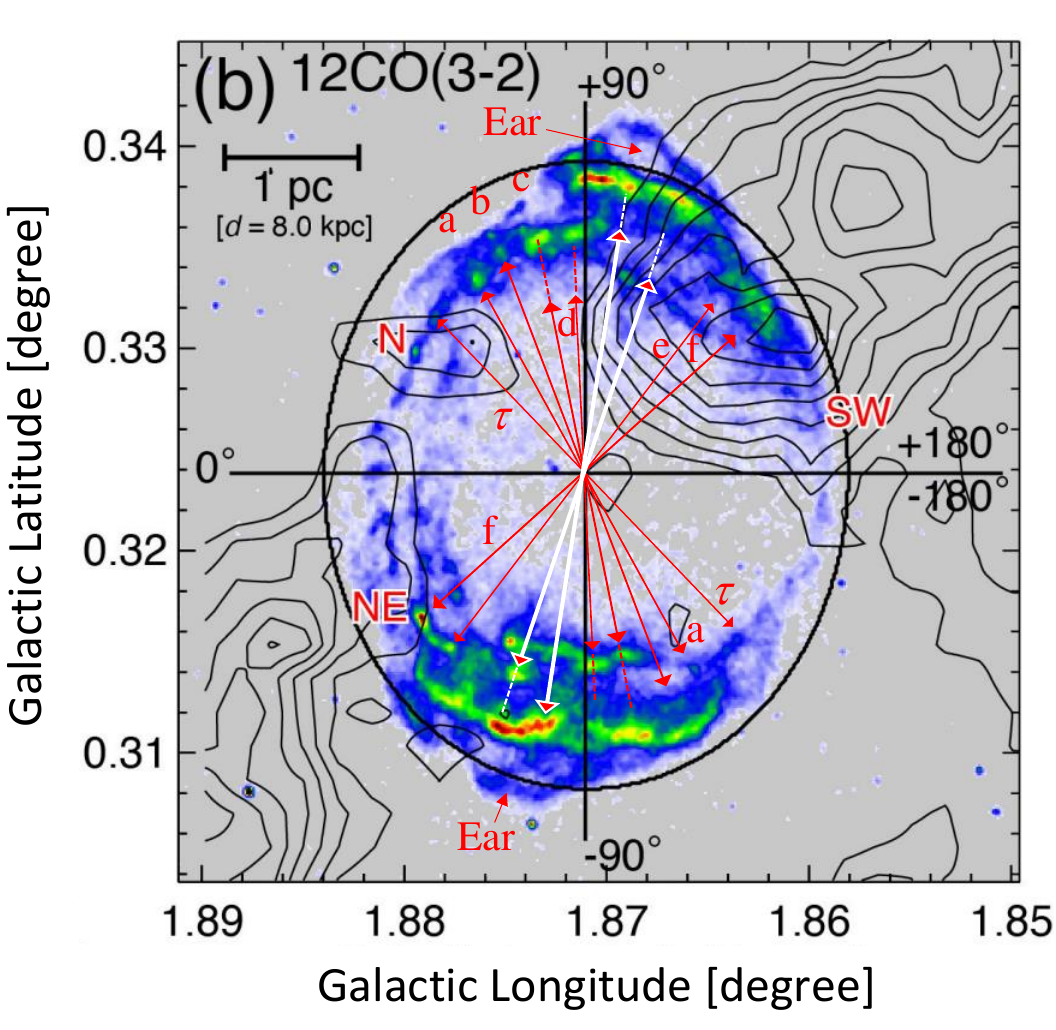}  
\includegraphics[trim=4.03cm 8.5cm 2.5cm 2.0cm ,clip, scale=0.4]{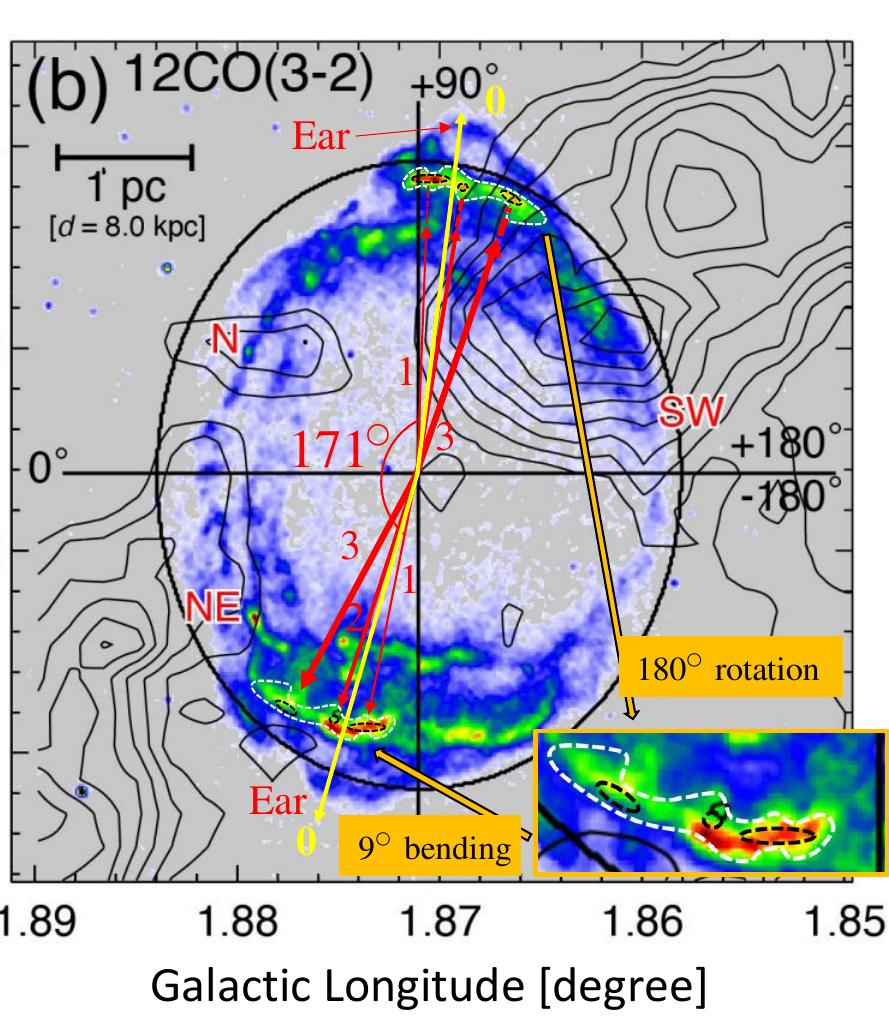}
\caption{An X-ray image image of SNR G1.9+0.3 with CO contours from \cite{Enokiyaetal2023}. The ellipse and the coordinate lines are in the original image. 
Left panel: an image with annotations from \cite{Soker2023SNRG1903} to present the point symmetric morphology: a mark of the two ears and the double-headed arrows (DHA), including dashed line continuations. DHA-a to DHA-f point at twin clumps of a point symmetric structure, while DHA-$\tau$ is a  tentative pair because the clumps are small and faint. The two white double-headed arrows point at pairs of clumps that require a different analysis as presented on the right panel (from \citealt{Soker2023SNRG1903}). The right panel includes the identification of the arc at the base of the upper (western) ear by a white dashed line, and its three peaks with dashed black lines. This is rotated around itself by $180^\circ$ and matched to the arc at the base of the bottom (eastern) ear, as the inset indicates. This procedure reveals a $9^\circ$ bend point symmetry of the two ears (DHA-0) and of the two base arcs (DHA-1 to DHA-3).
}
\label{Fig:SNRfig1}
\end{figure*}

\cite{Enokiyaetal2023} published a new X-ray image of the youngest supernova remnant (SNR) in the Galaxy, the type Ia SNR G1.9+0.3. In a recent paper \citep{Soker2023SNRG1903} I presented my identification of a point-symmetrical structure in this SNR Ia. Explosion models of SNe Ia do not form point-symmetrical ejecta. Therefore, I concluded that the point-symmetrical morphology is an imprint of the circumstellar material (CSM) into which the ejecta expanded since the explosion in about 1890-1900. Furthermore, the known substantial deceleration of the ejecta of SNR G1.9+0.3 (e.g., \citealt{Borkowskietal2017}) suggests a relatively massive CSM of $\gtrsim 1 M_\odot$. I therefore suggested that SNR G1.9+0.3 was an SN Ia inside a PN (SNIP). 
I present the identification of the point-symmetry in the two panels of Figure \ref{Fig:SNRfig1} (see caption for more details).  
% FFFFFFFFFFFFFFFFFFFFFFFFFFFFFFFFFFFFFFFFFFF  
%FFFFFFFFFFFFFFFFFFFFFFFFFFFFFFFFFFFFFFFFFF

The most likely scenario to account for such a SNIP is the core degenerate scenario, where a WD companion spiralling inside the AGB progenitor merges with the core at the end of the common envelope. In this case, there is a PN with a massive central star that is a merger product. The explosion of SNR G1.9+0.3 took place within $\lesssim 10^5 \yr$ after the common envelope and the formation of the PN. Another possible, but less likely, scenario for such a SNIP is the double degenerate scenario in which the WD companion ended the common envelope at a very close orbit to the core. Gravitational wave radiation brought the WD-core system to interact and explode within $\lesssim 10^5 \yr$. The PN in this case was ionized by the core and the WD companion. 

The point symmetry extends over a large angle from the double-headed arrow (DHA) line DHA-$\tau$ to line DHA-f (left panel). The two double ear structure (lines DHA-0 to DHA-3 on the right panel) has a bend of $9^\circ$. The pairs of clumps to the two sides of the ears do not share this bend. I, therefore, speculate that the PN progenitor of SNR G1.9+0.3 was a multipolar PN with some small departures from pure point symmetry. Due to the required massive nebula (as inferred from the decelerated SNR ejecta) and the ionizing source, which was a massive WD or a binary system of two WDs (as discussed above), I further speculate that this might have been a PN at or near the bright cutoff of the PNLF. I note that \cite{YaoQuataert2023} discuss the WD-WD merger as a scenario to form a bright hydrogen-poor PN (although no such bright PN is known). However, they do not consider the core degenerate scenario evolution through a regular (hydrogen-rich) PN.  

% ====================================
\section{Discussion and Summary}
\label{sec:Summary}
% ====================================

In this Proceedings paper I presented a preliminary attempt to try and learn about the brightest PNe from their morphologies. Because morphology analysis requires good resolution, this study deals only with Galactic PNe. Examining the morphologies of the brightest Galactic PNe, I identified that they tend to be multipolar with small to moderate departures from pure point symmetry (Section \ref{sec:features}). These lead to the first three conclusions below. Adding to these conclusions my very recent claim (\citealt{Soker2023SNRG1903}) that the youngest Galactic supernova, SNR G1.9+0.3,  was an SN Ia inside a PN (SNIP) and my conclusion in Section \ref{sec:G1903} above that this PN was multipolar, bring me also to speculate on the fourth and fifth conclusions below. 

The main conclusions are as follows. 
\begin{itemize}
    \item (1) The brightest PNe progenitors tend to experience strong (violent) binary interaction. This rules out a low-mass companion. I estimate that the companion has a mass of $M_2 \gtrsim 0.5 M_\odot$, which implies a main sequence or a WD companion. 
    \item (2) They are generally not the result of a triple stellar interaction. Namely, a tertiary star does not play a role in explaining the bright cutoff of the PNLF.  
    \item (3) Theoretical studies of the universal bright cutoff of the PNLF cannot ignore binary interaction. 
    \item (4) Consider the claim that SNR G1.9+0.3 was an SN Ia inside a multipolar PN, as I discussed in Section \ref{sec:G1903}. The likely scenario is one where a WD companion merged with the core of the AGB star at the end of the common envelope (the core degenerate scenario), or that they came very close to each other (the double degenerate scenario). The ionizing source of the pre-explosion PN progenitor of SNR G1.9+0.3 was either a massive WD that is a merger remnant, or a system of two close WDs. Theoretical studies should take into account the possibilities of such ionizing sources of the brightest PNe. 
    \item (5) I further speculate that some (but not all) of the central star of the brightest PNe are the merger product of a WD companion with the core. Some, but not all, of these, might explode as SNe Ia, possibly while the nebula still exists, i.e., SNIPs. 
\end{itemize}
$$ $$
\textbf{Acknowledgement.}
% ===========================================
I thank Nick Chornay for the list of the brightest PNe. In this research, I used the \textit{PNIC: Planetary Nebula Image Catalogue} composed by Bruce Balick (http://faculty.washington.edu/balick/PNIC/). I acknowledge funding from the Pazy Research Foundation in Israel. 

% BBBBBBBBBBBBBBBBBBBBBBBBBBBBBBBBBBBBBBB

\end{document}